\documentclass[aps,prl,superscriptaddress,twocolumn]{revtex4-2}
\usepackage{amsmath}  
\usepackage{amsfonts} 
\usepackage{graphicx} 
\usepackage{url}
\usepackage{natbib}

\begin{document}
\title{Boundary Condition Induced Passive Chaotic Mixing in Straight Microchannels}
\author{Habilou Ouro-Koura}
\affiliation{Department of Mechanical Engineering,
Rensselaer Polytechnic Institute, Troy, NY 12180 USA}
\affiliation{University of Maryland Eastern Shore, 1, Backbone Road, Princess Anne, MD 21853 USA}
\author{Ayobami Ogunmolasuyi}
\affiliation{Department of Mechanical Engineering,
Dartmouth College, Troy, NY 12180 USA}
\affiliation{University of Maryland Eastern Shore, 1, Backbone Road, Princess Anne, MD 21853 USA}
\author{Othman Suleiman}
\author{Jaylah Easter}
\affiliation{University of Maryland Eastern Shore, 1, Backbone Road, Princess Anne, MD 21853 USA}
\author{Yasmin Roye}
\affiliation{Department of Biomedical Engineering, Pratt School of Engineering, Duke University, Durham, NC 27708, USA}
\affiliation{University of Maryland Eastern Shore, 1, Backbone Road, Princess Anne, MD 21853 USA}
\author{Othman Suleiman}
\affiliation{University of Maryland Eastern Shore, 1, Backbone Road, Princess Anne, MD 21853 USA}
\author{Kausik S Das}
\email{kdas@umes.edu} 
\affiliation{University of Maryland Eastern Shore, 1, Backbone Road, Princess Anne, MD 21853 USA}

\date{\today}
\begin{abstract}
 Mixing in low Reynolds number flow is difficult because in this laminar regime it occurs mostly via slow molecular diffusion. This letter reports a simple way to significantly enhance low Reynolds number passive microfluidic flow mixing in a straight microchannel by introducing asymmetric wetting boundary conditions on the floor of the channel. We show experimentally and numerically that by creating carefully chosen hydrophobic slip patterns on the floor of the channels we can introduce stretching, folding and/or recirculation in the flowing fluid volume, the essential elements to achieve mixing in absence of turbulence. We also show that there are two distinctive pathways to produce homogeneous mixing in microchannels induced by the inhomogeneity of the boundary conditions. It can be achieved either by: 1) introducing stretching, folding and twisting of fluid volumes, i.e., via a horse-shoe type transformation map, or 2) by creating chaotic advection, through manipulation of the hydrophobic boundary patterns on the floor of the channels. We have also shown that by superposing stretching and folding with chaotic advection, mixing can be optimized by significantly reducing mixing length, thereby opening up new design opportunities for simple yet efficient passive microfluidic reactors.

\end{abstract}
\maketitle

Microfluidics deals with control and manipulation of small volumes of fluids through channels with characteristic length scales in the micron order an it exercises precise dynamic control over the flow to study new phenomena occurring in fluids. Microfluidics has the potential to solve some of the grand engineering challenges such as engineering better medicines, providing access to clean water, solving energy problems due to its ability to use very small controlled volume of samples and reagents, high resolution separation and detection with great sensitivity\cite{whitesides2006origins, fraikin2011high, sinton2014energy} etc. Low manufacturing cost, short time scales for analysis along with high throughput designs using device miniaturization such as lab-on-a-chip\cite{mark2010microfluidic} or a reactor-on-a-chip\cite{srisa2008fluorescence} are other advantages. The application of microfluidics is manyfold, ranging from microelectronics, bioanalysis\cite{sackmann2014present, martinez2009diagnostics}, nanoparticle synthesis, organ in a chip, drug development, testing and controlling multiphase flow\cite{teh2008droplet, ando2012homogeneous, nordstrom2010microfluidic}, optofluidics, acoustofluidic patterning\cite{collins2018self}, to detection of a single cell \cite{nagrath2007isolation, burg2007weighing}, single file diffusion\cite{locatelli2016single} and even a single molecule\cite{dittrich2005single}.

It is clear that microfluidics offers solutions to a plethora of problems, however, growth of microfluidic technlogies to the desired level has been stalled by some bottlenecks.  One major bottleneck in developing an efficient microfluidic device is to mix fluids in a microchannel passively, where miscible fluid components can be mixed homogeneously and spontaneously in low Raynolds number regime. Most of the applications of microfluidic chips involving chemical and biological components homogeneous mixing of reactants is essential to achieve uniform solution environments. However, it is difficult to mix component fluids in microchannels spontaneously, thanks to the absence of turbulence in the viscosity dominated laminar flow regimes in the small characteristic length scales. Microfluidic mixers are thus of pivotal importance for efficient functioning of microfluidic devices for a range of important applications.

In a typical microchannel cross-section length scale (\textit{l}) is $\sim 100 $ micron or less. With small flow rates $U\sim0.1$m/s, Reynolds numbers ($\textrm{Re}=Ul\rho/\mu$) are typically of the order of 10 or less, where $\mu$ is the dynamic viscosity ($\sim10^{-3}$ Pa.S) and $\rho$ is the density ($\sim10^3$ Kg/m$^{3}$ for water) of the fluid. This means that not only inertia is negligible, but also molecular diffusion is the dominant mechanism of mixing of the component fluids in this regime.
However, typical diffusivity ($D$) of species varies from approximately 10$^{-9}$ \textrm{m$^2$/s}  for small molecules including ions, to 10$^{-11}$ \textrm{m$^2$/s} for large biomolecules.
As a result, mixing time scale ($l^2/D >$10 s) and P\'{e}clet number ($Pe=Ul/D$), the relative measure of diffusive time scale over advective time scale, are large (10$^4$ $-$ 10$^6$). The length required to mix via diffusion in a flow thus ($l_{mix}=lPe$) typically ranges between tens of centimeters to even in meters!

 One of the strategies that can be employed to enhance mixing in steady laminar diffusion dominated flow regimes comes through a Smale horseshoe map and/or a baker's map \cite{christov2011stretching}. By effectively decreasing the striation length, $l_{\rm Str}\rightarrow 0$, defined as the length scale over which diffusion acts to homogenize the concentration, these maps transform the space (the flow domain) into itself through successive stretching and folding. The effective interface area across which diffusion takes place thus increases exponentially leading to a positive Lyapunov exponent. In other words this is a route to chaos or a rapid growth in mixing, when the problem is translated into the language of dynamical systems.

 The other pathway to achieve mixing in low Reynolds number flow goes through `chaotic advection', where the fluid still associates itself with stretching and folding on top of switching flows instantaneously from one streamline to another\cite{wiggins2004foundations}. This mechanism of mixing is observed in blinking flows and can be understood theoretically using a twist map\cite{aref1984stirring}. Although both the transformations can be analyzed rigorously using mathematical concepts of ergodicity and dynamical systems, a practical demonstration of these idealized mathematical models is rather difficult to achieve. Mixing via chaotic advection was first  demonstrated in the pioneering work on the Staggered Herringbone Mixer (SHM) \cite{stroock2002chaotic}, where flow transverse to the primary flow direction suitable for chaotic advection was generated by placing three dimensional patterned ridges on the floor of the channels at oblique angles with respect to the downstream direction of the flow. Mixing length varied linearly with $\textrm{ln(Pe)}$ confirming that the stretching and folding of the volumes grow exponentially with respect to the effective mixing length. In this paper we take a totally new approach to generate similar transverse flow by creating anisotropic wetting conditions on the floor of the channels experimentally and numerically. Moreover, we show that this type of mixing is not limited to just Harringbone type patterns, rather a range of  CFD simulations reveal the effects of many different patterns on the flow behavior and their corresponding mixing mechanisms in simple microfluidic channels.

 It is well known that surfaces play a big role in micro and nanoscales because the boundary layer essentially spans the whole cross sections of the channels. Since the ratio of solid-liquid interface area ($\textrm{A}\sim l^2$) in a channel and the volume of fluid ($\textrm{V}\sim l^3$) varies inversely proportional to the characteristic length scale ($\sim 1/l $) of the system, in small length scales such as micro or nano channels ($l\rightarrow 0$), the effect of boundary conditions become increasingly dominant in this asymptotic regime. In this work we exploit this fact and report that unlike \cite{stroock2002chaotic},  where 3D miropatterning involving rather complex fabrication process was needed to achieve chaotic advection, here we have achieved mixing in a microchannel at Re $\leq$10 by creating hyrdophobic patterns on the floor using water-repellant ink dots on the channel floor. To be precise, we have used simple hydrophobic permanent marker ink\cite{khodaparast2017water} that contains hydrophobic polymer resin to create these hydrophobic patterns on one of the inner walls of the straight channel.  These hydrophobic dots create a discontinuous change in the wall boundary conditions forcing the streamlines to readjust instantaneously while crossing the boundary between no-slip to hydrophobic regions.  Obviously there is less resistance to flow over the water-repellant regions and the liquid flows faster over them than the no-slip regions on the floor of the channels. Moreover, the hydrophobic patterns constrain the fluid to avoid those regions, in turn potentially forcing the fluid to generate a velocity component in the transverse direction. We hypothesize that because of this anisotropic and abrupt flow resistance change due to the difference in wetting boundary conditions on the floor, the fluid generates an average transverse flow, resulting in twisting and stretching of the fluid volume over the cross section of the channel, or creating counter rotating recirculation zones leading to chaotic advection.

  \begin{figure}[ht!]
\centering
\includegraphics[angle=0,width=3.4in]{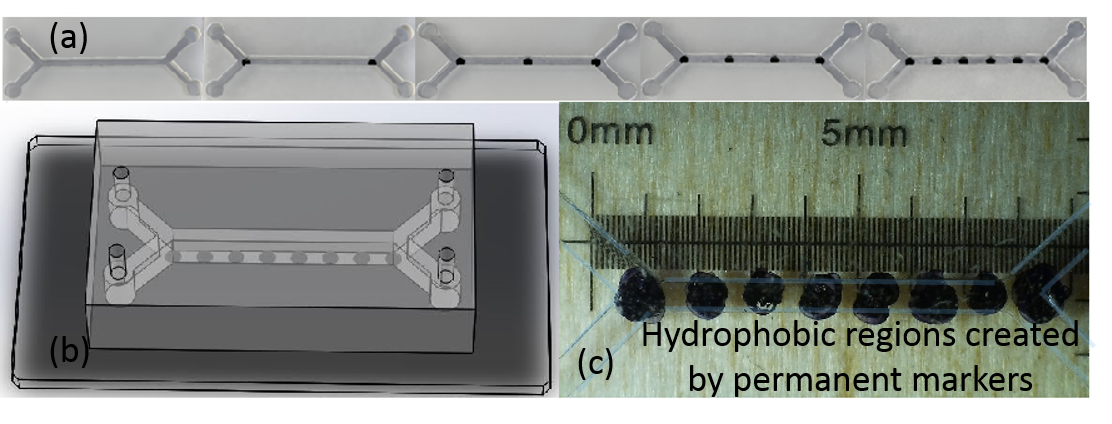}
\caption{Design of the straight microchannels with asymmetric boundary condition. Dark spots are the hydrophobic regions on the floor of the microchannels.}
\label{schematic2bcfractal}
\end{figure}

The whole fabrication process of microchannels with anisotropic boundary conditions is quite simple and can be done using soft photolithography process with polydimethylsiloxane (PDMS) elastomer. Designs of various microchannels are shown in Fig.\ref{schematic2bcfractal}. To create slip regions on the floor of the channels we have used an unique technique. Recently, it has been shown that many commercially available permanent marker inks, such as Sharpie\textregistered  \; inks create elastic thin polymer films upon drying on a substrate and exhibit water-repellent hydrophobic behavior \cite{khodaparast2017water}. Taking advantage of this property and easy availability of school-supply permanent markers, we have created slip dots on the floor of the channels by simply placing the dots using an ultrafine Sharpie\textregistered  \;marker manually. After the hydrophobic dots are created on a glass slide (Fig.\ref{schematic2bcfractal}) PDMS open channels created by standard soft photolithography process\cite{xia1998soft} are bonded with the slide using plasma boding techniques\cite{barnes2021plasma, bhattacharya2005studies}.  The hydrophobic spots created manually on the floor of the channels are not fully symmetrical as expected, and can be seen in Fig.\ref{schematic2bcfractal} c. Several microchannels are made with width and depth in the range of 100$\mu$m to 300 $\mu$m, with the length of the channels 5-10 mm. In the experiments Reynolds number is always kept at 10 by controlling the flow speed in the syringe pumps.

 \begin{figure}[ht!]
\centering
\includegraphics[angle=0,width=3.4in]{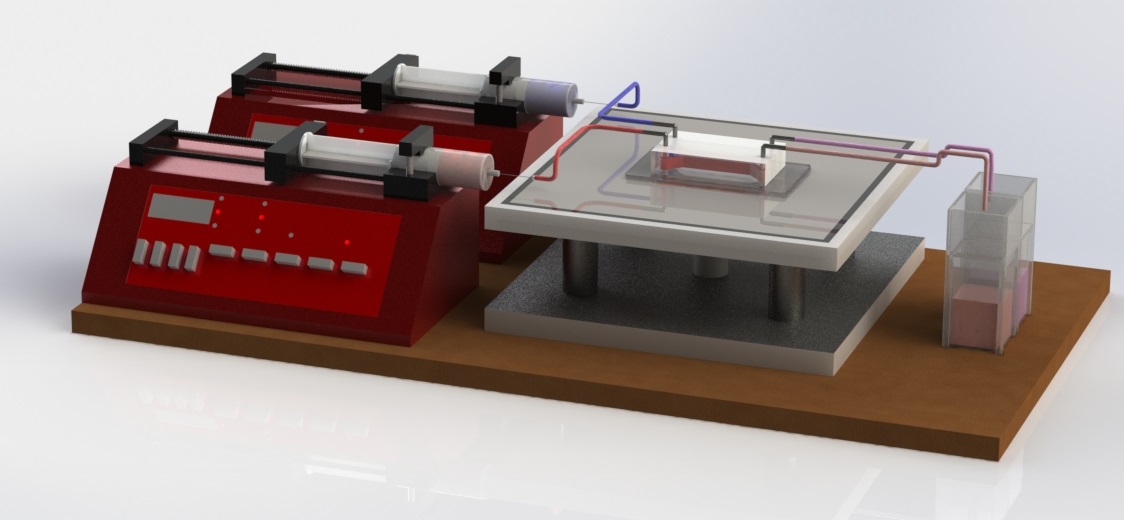}
\caption{Schematic of the microfluidic mixing experiment. One of the syringe pumps pushes a colored fluid and the other brings plain water to the microfluidic chip. Fluid is collected at the outlets  after going through the mixer and analyzed by a UV-VIS spectrometers.}
\label{schematic1pump}
\end{figure}

Fig.\ref{schematic1pump} shows the schematic of our experiments where two miscible fluids are pumped into the microfluidic reactor through the inlets by two syringe pumps and collected in the cuvettes placed at the outlet. The fluids collected at the outlet are analyzed by UV-VIS absorption spectroscopy to determine the mixing index and the overall effectiveness of the hydrophobic spots in the mixing of component fluids. In our experiment we use deionized (DI) distilled water (Barnstead NANOpureDiamond water purification system, specific resistivity 18.2 M$\Omega$-cm) in one of the syringe pumps, and DI water mixed with Allura Red dye (D=3.23$\times10^{-10}$m$^2$/s) in the other. In a simple microchannels with usual no-slip inner walls the output confirms a fully separated flow, i.e, one of the cuvettes collects red colored fluid with same concentration of the input colored fluid, confirmed by the UV-VIS spectroscopic analysis, whereas the other output cuvette collects colorless water. This fully separated flow is characterized by the mixing index 0.5. On the other hand as the hydrophobic marker dots are introduced, two input liquids start to mix. A fully mixed homogeneous fluid's concentration is analyzed by the spectrophotometer and the mixing index is scaled as 0.



\begin{figure}[ht!]
\centering
\includegraphics[angle=0,width=3.4in]{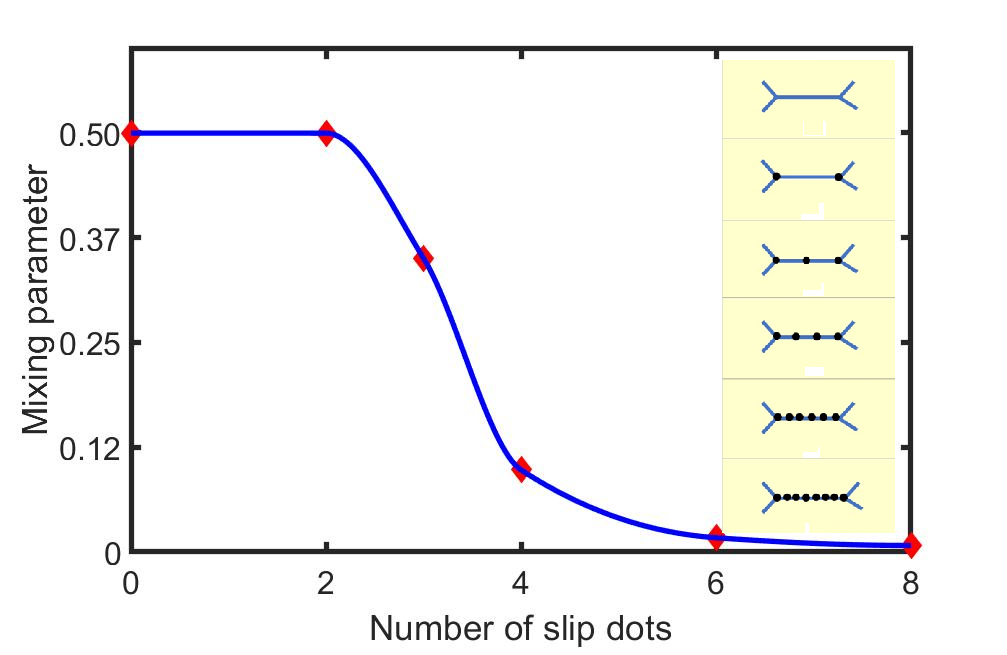}
\caption{Experimental results on the effect of periodic hydrophobic regions on mixing in a straight 7mm microchannel. Normalized mixing index 0.5 signifies separated flow and mixing index 0 signifies homogeneous mixing.}
\label{slip_straight}
\end{figure}
 The effect of the slip dots on the degree of mixing at the outlet of the channels, measured experimentally using a spectrophotometer is shown in Fig.\ref{slip_straight}. A transition from laminar separated flow to a homogeneous mixing of component fluids is observed when the number of hydrophobic dots exceeds 3 in a 7mm channel.

To understand the details of the mixing mechanism a numerical investigation is done using the finite element simulation software package COMSOL. Navier-Stokes and convection diffusion equations are solved with hydrophobic spots considered as slip regions in the boundary conditions, while imposing no slip conditions on the rest of the inner wall regions of the channel. Steady incompressible flow with laminar outflow and zero outlet static pressure is employed. In the code Reynolds number is kept at 10 to begin with, to match it with our experimental conditions.

\begin{figure}[ht!]
\centering
\includegraphics[angle=0,width=3.4in]{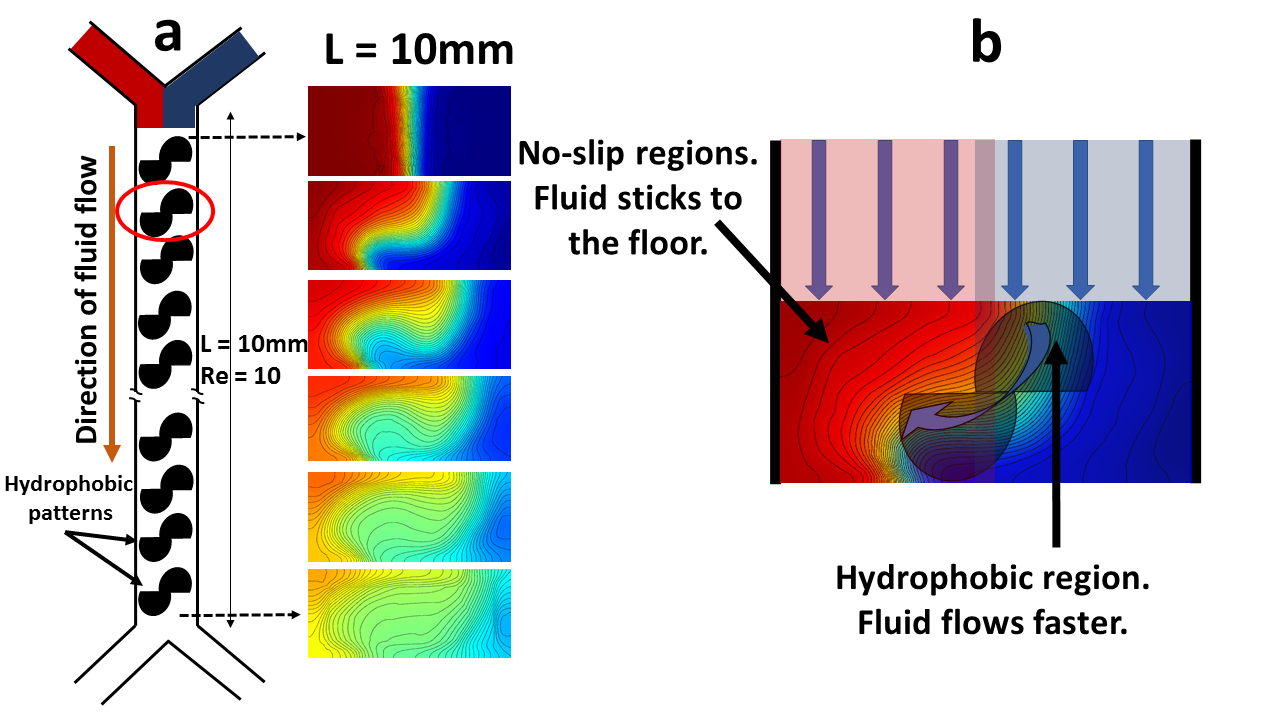}
\caption{Effect of periodic hydrophobic regions on mixing in a straight microchannel. a) The asymmetric duck-shape hydrophobic regions introduce difference of resistance to flow about the centerline of the channel. b) A zoomed version of one of the hydrophobic regions shows the possible path of the fluid that flows over the the shapes. It is clear that this pattern forces the fluids to bends the fluid and one component of the fluid intrudes the other thereby introducing effective transverse rotation and stretching and folding of the fluid volume as it flows downstream.}
\label{Duck}
\end{figure}

To further benchmark the code we first created herringbone type 2D hydrophobic patters on the floor of the channels to mimic the 3D grooved channels in SHM previously reported\cite{stroock2002chaotic}. It shows similar mixing profiles (Fig. 3 of \cite{stroock2002chaotic}) in comparison to our simulation results (Fig.\ref{Herringbone} of this paper). From the numerical simulation we have identified two different pathways to achieve homogeneous mixing of component fluids in the low Reynolds number laminar flow regime. Preferential bending of the flow can be achieved by introducing periodic asymmetries in the boundary conditions, created by asymmetric hydrophobic patterns about the centerline of the channel shown in Fig.\ref{Duck}a. Analogous to Magnus effect\cite{barkla1971magnus}, where difference between the speeds of a fluid flowing over two opposite sides of an object gives rise to a pressure difference in the transverse direction results in bending, here our `duck' shape patterns, for example, create a transverse pressure gradient generated through the difference of uneven flow resistances about the centerline.

We argue that the floor of the channels is no-slip everywhere except the locations where there is a hydrophobic film. So, the hydrophobic regions on the floor applies less resistance to the fluid flow in comparison to the corresponding no slip region on the other side of the centerline. Each of the patterns thus may direct one component fluid (in our case could be blue) to intrude into the other (red), eventually creating a twisting, stretching and folding of both the components as the fluids flow downstream and leading to a Smale horshoe type mixing pathway as seen in Fig.\ref{Duck}.

 \begin{figure}[ht!]
\centering
\includegraphics[angle=0,width=3.4in]{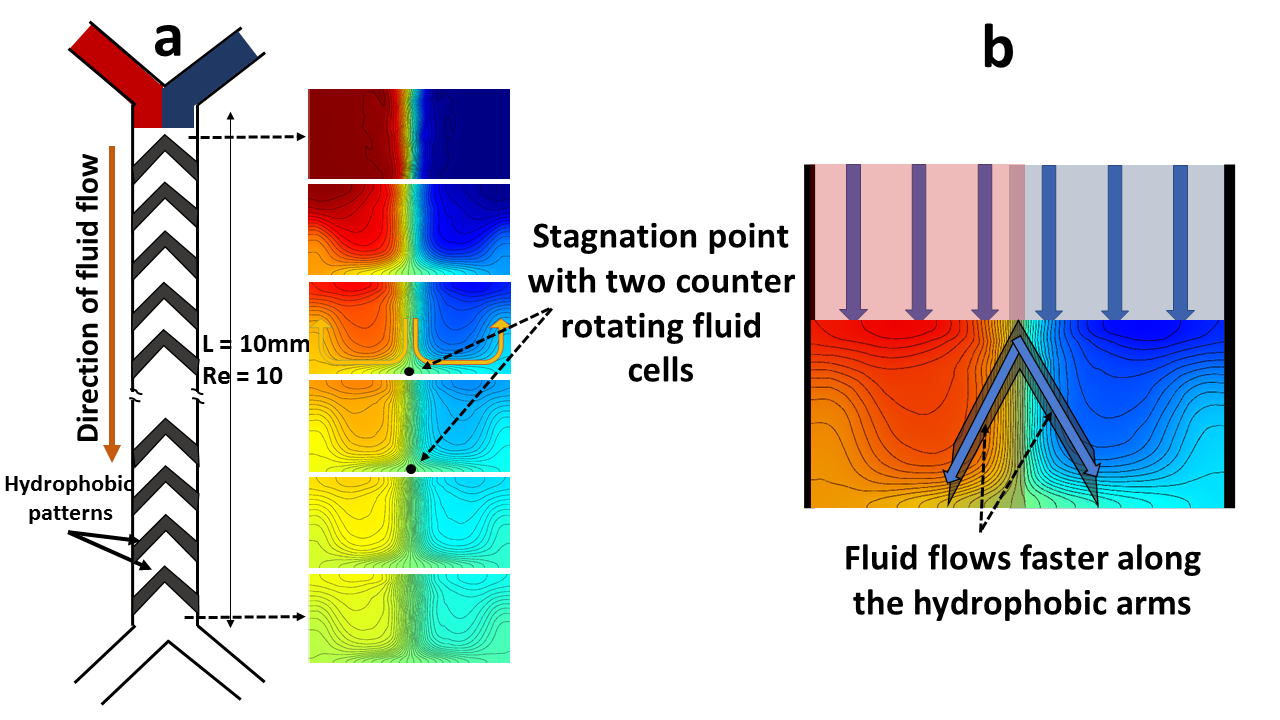}
\caption{Effect of periodic hydrophobic regions on mixing in a straight microchannel through symmetric V shapes hydrophobic patches. a) The symmetric V shaped regions introduce difference of resistance to flow along the arms. b) A zoomed version of one of the hydrophobic regions shows the path of the fluid that flows along the arms. An important observation is that the vertex of the V shape divides the flow and channel them obliquely in two opposite directions, thus creating counter rotating recirculation zones.}
\label{Vshape}
\end{figure}

Another pathway of mixing is found by creating counter rotating recirculation zones with the help of `V' shaped hydrophobic patterns (Fig. \ref{Vshape} ). It is clear that the stagnation point between the recirculating fluids align with the position of the apex point of the `V' shape. This feature is evident in the herringbone pattern (Fig. \ref{Herringbone}) also, where the stagnation points can be shifted periodically in the transverse direction by moving the apex points of the asymmetric `V' shapes. This creates different recirculation patterns as the fluids flow downstream leading towards effective mixing through chaotic advection. To check if this enhanced mixing is an artefact of chaos we investigated the power-law dependence of the mixing length on the  P\'{e}clet number $Pe$. When mixing occurs only via diffusion, the channel length required for mixing is given by $l_{mix}\sim Ut_{diff}\sim Ul^2/D\sim lPe$, whereas if there is indeed chaotic root to mixing via stretching and folding of  liquid volumes over the cross section of the channel then the mixing length would vary as  $l_{mix}\sim l\textrm{ ln}(Pe)$ \cite{ottino1989kinematics, stroock2002chaotic, jones1989chaotic}. A linear profile of $l_{mix}$ with respect to $\textrm{ln}(Pe)$ can thus be considered as a signature of chaotic advection in the system. In Fig.{\ref{Peclet} we have indeed seen this behavior in our irregular V shaped passive microfluidic mixers.

 \begin{figure}[ht!]
\centering
\includegraphics[angle=0,width=3.4in]{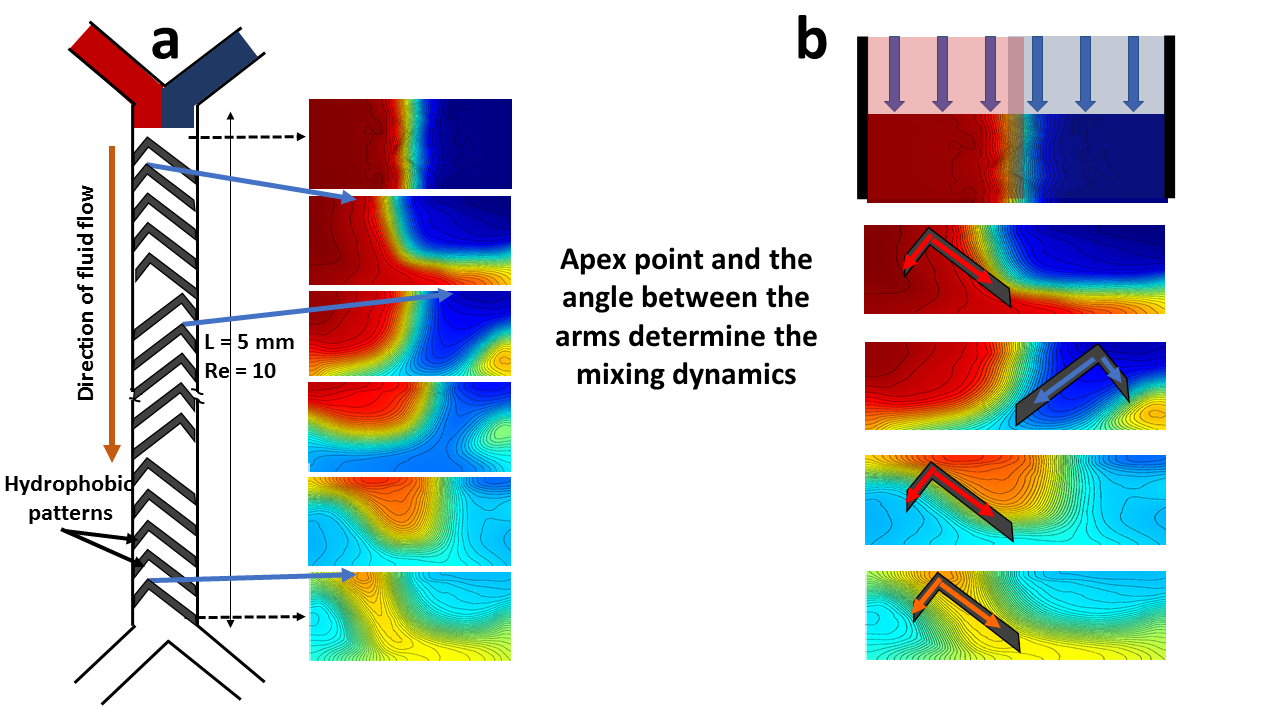}
\caption{Effect of periodic hydrophobic regions on mixing in a straight microchannel. a) The asymmetric herringbone shaped hydrophobic regions introduce difference of resistance to flow along the arms of the asymmetric V shapes. b) A zoomed version of one of these regions shows the path of the fluid that flows over the the patches. One can think of the Harrinbone structures as irregular V shapes. Here also the vertices introduce recirculation. However, as the vertices change location depending on the wavelength of the pattern, stagnation points of the recirculations also move back and forth, in turn introducing chaotic advection in the flow.}
\label{Herringbone}
\end{figure}

 \begin{figure}[ht!]
\centering
\includegraphics[angle=0,width=3.5in]{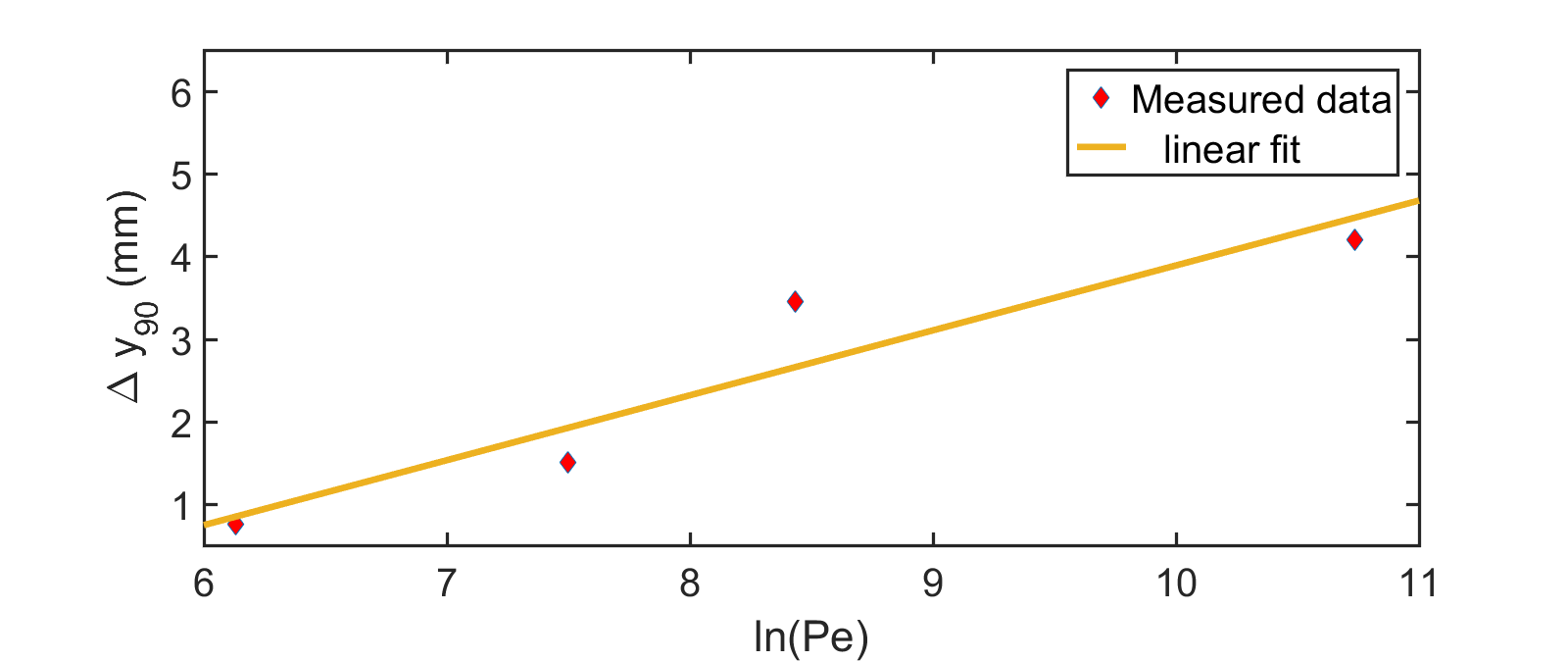}
\caption{Effect of Peclet number on mixing in a Herringbone mixer. A linear relationship between $\textrm{ln(Pe)}$ and the mixing length, where 90\% of mixing occurs indicates route to chaos\cite{stroock2002chaotic} in a low Re environment.}
\label{Peclet}
\end{figure}

We were also curious to see if we can create a superposition of the Magnus type bending with the recirculating cells using both the pathways mentioned above and the results are shown in Fig. \ref{Superposition}. It shows that alternating twisting and recirculation gives rise to excellent mixing in low Reynolds number laminar flow in a straight microchannel.

\begin{figure}[ht!]
\centering
\includegraphics[angle=0,width=3.4in]{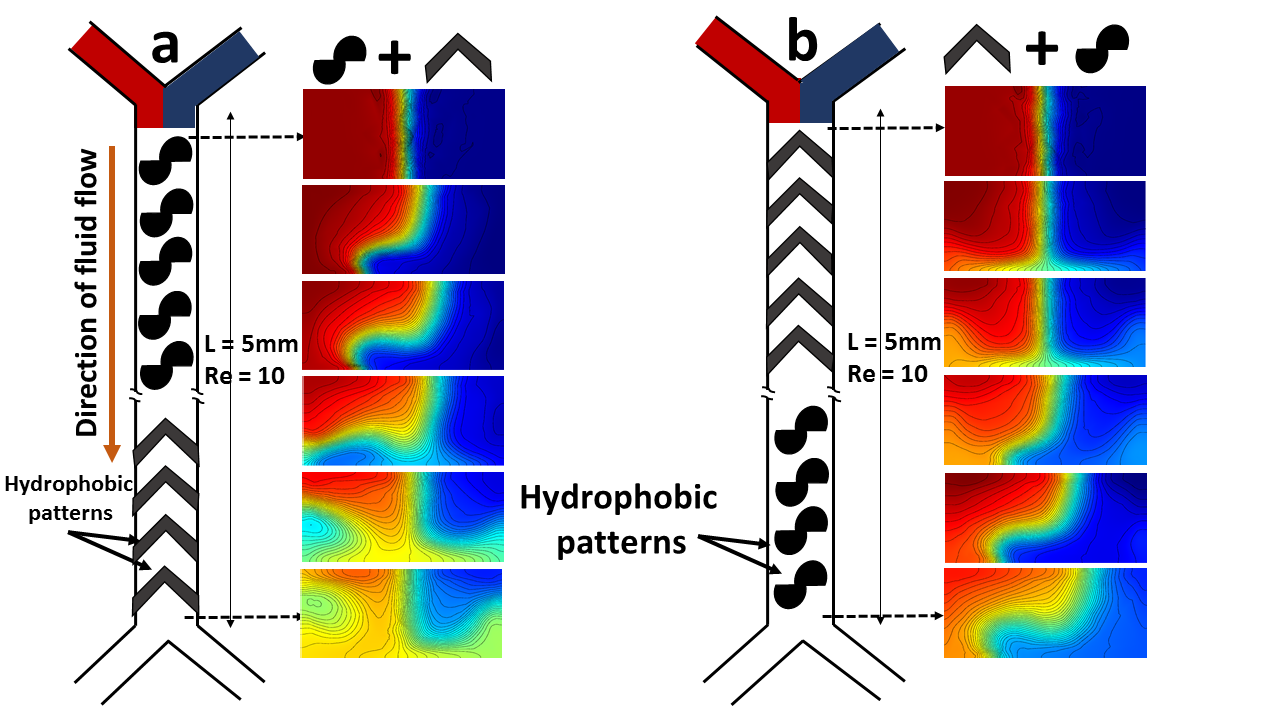}
\caption{Effect of periodic hydrophobic regions on mixing in a straight microchannel through hybrid patterns. We have observed a superposition of effects of stretching through rotation (by duck shapes) and recirculation (by V shapes) or vice versa. a) Here the duck shape patterns were created first close to the inlet followed by V shape patterns. It is clear that one component of the fluid intrudes into the other and introduces stretching and folding by the duck shape as discussed earlier.  On top of that the symmetric V shapes create recirculating flow patterns leading towards complex mixing effect. b) An opposite effect is observed when the pattern is switched, i.e., V shapes followed by the duck shapes. Recirculation followed by stretching and folding give rise to a different mixed state.}
\label{Superposition}
\end{figure}

In conclusion we claim that rapid mixing in low Reynolds number flow even in a straight microchannel can be achieved by manipulating twisting, stretching and recirculation in the flow created by passive interventions through carefully chosen hydrophobic patterns on the walls/floor of the microchannels. In this case we won't even need complex three dimensional micro/nano patterning on the floor of the channels\cite{stroock2002chaotic} to introduce chaos in this regime. This letter may lead the microfluidic industry to an easy and inexpensive fabrication process to achieve efficient homogeneous mixing with a small mixing length scale.




\bibliography{BibtexdatabaseKSD}

\begin{thebibliography}{26}%
\makeatletter
\providecommand \@ifxundefined [1]{%
 \@ifx{#1\undefined}
}%
\providecommand \@ifnum [1]{%
 \ifnum #1\expandafter \@firstoftwo
 \else \expandafter \@secondoftwo
 \fi
}%
\providecommand \@ifx [1]{%
 \ifx #1\expandafter \@firstoftwo
 \else \expandafter \@secondoftwo
 \fi
}%
\providecommand \natexlab [1]{#1}%
\providecommand \enquote  [1]{``#1''}%
\providecommand \bibnamefont  [1]{#1}%
\providecommand \bibfnamefont [1]{#1}%
\providecommand \citenamefont [1]{#1}%
\providecommand \href@noop [0]{\@secondoftwo}%
\providecommand \href [0]{\begingroup \@sanitize@url \@href}%
\providecommand \@href[1]{\@@startlink{#1}\@@href}%
\providecommand \@@href[1]{\endgroup#1\@@endlink}%
\providecommand \@sanitize@url [0]{\catcode `\\12\catcode `\$12\catcode
  `\&12\catcode `\#12\catcode `\^12\catcode `\_12\catcode `\%12\relax}%
\providecommand \@@startlink[1]{}%
\providecommand \@@endlink[0]{}%
\providecommand \url  [0]{\begingroup\@sanitize@url \@url }%
\providecommand \@url [1]{\endgroup\@href {#1}{\urlprefix }}%
\providecommand \urlprefix  [0]{URL }%
\providecommand \Eprint [0]{\href }%
\providecommand \doibase [0]{https://doi.org/}%
\providecommand \selectlanguage [0]{\@gobble}%
\providecommand \bibinfo  [0]{\@secondoftwo}%
\providecommand \bibfield  [0]{\@secondoftwo}%
\providecommand \translation [1]{[#1]}%
\providecommand \BibitemOpen [0]{}%
\providecommand \bibitemStop [0]{}%
\providecommand \bibitemNoStop [0]{.\EOS\space}%
\providecommand \EOS [0]{\spacefactor3000\relax}%
\providecommand \BibitemShut  [1]{\csname bibitem#1\endcsname}%
\let\auto@bib@innerbib\@empty
\bibitem [{\citenamefont {Whitesides}(2006)}]{whitesides2006origins}%
  \BibitemOpen
  \bibfield  {author} {\bibinfo {author} {\bibfnamefont {G.~M.}\ \bibnamefont
  {Whitesides}},\ }\bibfield  {title} {\bibinfo {title} {The origins and the
  future of microfluidics},\ }\href@noop {} {\bibfield  {journal} {\bibinfo
  {journal} {Nature}\ }\textbf {\bibinfo {volume} {442}},\ \bibinfo {pages}
  {368} (\bibinfo {year} {2006})}\BibitemShut {NoStop}%
\bibitem [{\citenamefont {Fraikin}\ \emph {et~al.}(2011)\citenamefont
  {Fraikin}, \citenamefont {Teesalu}, \citenamefont {McKenney}, \citenamefont
  {Ruoslahti},\ and\ \citenamefont {Cleland}}]{fraikin2011high}%
  \BibitemOpen
  \bibfield  {author} {\bibinfo {author} {\bibfnamefont {J.-L.}\ \bibnamefont
  {Fraikin}}, \bibinfo {author} {\bibfnamefont {T.}~\bibnamefont {Teesalu}},
  \bibinfo {author} {\bibfnamefont {C.~M.}\ \bibnamefont {McKenney}}, \bibinfo
  {author} {\bibfnamefont {E.}~\bibnamefont {Ruoslahti}},\ and\ \bibinfo
  {author} {\bibfnamefont {A.~N.}\ \bibnamefont {Cleland}},\ }\bibfield
  {title} {\bibinfo {title} {A high-throughput label-free nanoparticle
  analyser},\ }\href@noop {} {\bibfield  {journal} {\bibinfo  {journal} {Nature
  nanotechnology}\ }\textbf {\bibinfo {volume} {6}},\ \bibinfo {pages} {308}
  (\bibinfo {year} {2011})}\BibitemShut {NoStop}%
\bibitem [{\citenamefont {Sinton}(2014)}]{sinton2014energy}%
  \BibitemOpen
  \bibfield  {author} {\bibinfo {author} {\bibfnamefont {D.}~\bibnamefont
  {Sinton}},\ }\bibfield  {title} {\bibinfo {title} {Energy: the microfluidic
  frontier},\ }\href@noop {} {\bibfield  {journal} {\bibinfo  {journal} {Lab on
  a Chip}\ }\textbf {\bibinfo {volume} {14}},\ \bibinfo {pages} {3127}
  (\bibinfo {year} {2014})}\BibitemShut {NoStop}%
\bibitem [{\citenamefont {Mark}\ \emph {et~al.}(2010)\citenamefont {Mark},
  \citenamefont {Haeberle}, \citenamefont {Roth}, \citenamefont {Von~Stetten},\
  and\ \citenamefont {Zengerle}}]{mark2010microfluidic}%
  \BibitemOpen
  \bibfield  {author} {\bibinfo {author} {\bibfnamefont {D.}~\bibnamefont
  {Mark}}, \bibinfo {author} {\bibfnamefont {S.}~\bibnamefont {Haeberle}},
  \bibinfo {author} {\bibfnamefont {G.}~\bibnamefont {Roth}}, \bibinfo {author}
  {\bibfnamefont {F.}~\bibnamefont {Von~Stetten}},\ and\ \bibinfo {author}
  {\bibfnamefont {R.}~\bibnamefont {Zengerle}},\ }\bibfield  {title} {\bibinfo
  {title} {Microfluidic lab-on-a-chip platforms: requirements, characteristics
  and applications},\ }in\ \href@noop {} {\emph {\bibinfo {booktitle}
  {Microfluidics Based Microsystems}}}\ (\bibinfo  {publisher} {Springer},\
  \bibinfo {year} {2010})\ pp.\ \bibinfo {pages} {305--376}\BibitemShut
  {NoStop}%
\bibitem [{\citenamefont {Srisa-Art}\ \emph {et~al.}(2008)\citenamefont
  {Srisa-Art}, \citenamefont {Demello},\ and\ \citenamefont
  {Edel}}]{srisa2008fluorescence}%
  \BibitemOpen
  \bibfield  {author} {\bibinfo {author} {\bibfnamefont {M.}~\bibnamefont
  {Srisa-Art}}, \bibinfo {author} {\bibfnamefont {A.~J.}\ \bibnamefont
  {Demello}},\ and\ \bibinfo {author} {\bibfnamefont {J.~B.}\ \bibnamefont
  {Edel}},\ }\bibfield  {title} {\bibinfo {title} {Fluorescence lifetime
  imaging of mixing dynamics in continuous-flow microdroplet reactors},\
  }\href@noop {} {\bibfield  {journal} {\bibinfo  {journal} {Physical Review
  Letters}\ }\textbf {\bibinfo {volume} {101}},\ \bibinfo {pages} {014502}
  (\bibinfo {year} {2008})}\BibitemShut {NoStop}%
\bibitem [{\citenamefont {Sackmann}\ \emph {et~al.}(2014)\citenamefont
  {Sackmann}, \citenamefont {Fulton},\ and\ \citenamefont
  {Beebe}}]{sackmann2014present}%
  \BibitemOpen
  \bibfield  {author} {\bibinfo {author} {\bibfnamefont {E.~K.}\ \bibnamefont
  {Sackmann}}, \bibinfo {author} {\bibfnamefont {A.~L.}\ \bibnamefont
  {Fulton}},\ and\ \bibinfo {author} {\bibfnamefont {D.~J.}\ \bibnamefont
  {Beebe}},\ }\bibfield  {title} {\bibinfo {title} {The present and future role
  of microfluidics in biomedical research},\ }\href@noop {} {\bibfield
  {journal} {\bibinfo  {journal} {Nature}\ }\textbf {\bibinfo {volume} {507}},\
  \bibinfo {pages} {181} (\bibinfo {year} {2014})}\BibitemShut {NoStop}%
\bibitem [{\citenamefont {Martinez}\ \emph {et~al.}(2010)\citenamefont
  {Martinez}, \citenamefont {Phillips}, \citenamefont {Whitesides},\ and\
  \citenamefont {Carrilho}}]{martinez2009diagnostics}%
  \BibitemOpen
  \bibfield  {author} {\bibinfo {author} {\bibfnamefont {A.~W.}\ \bibnamefont
  {Martinez}}, \bibinfo {author} {\bibfnamefont {S.~T.}\ \bibnamefont
  {Phillips}}, \bibinfo {author} {\bibfnamefont {G.~M.}\ \bibnamefont
  {Whitesides}},\ and\ \bibinfo {author} {\bibfnamefont {E.}~\bibnamefont
  {Carrilho}},\ }\bibfield  {title} {\bibinfo {title} {Diagnostics for the
  developing world: Microfluidic paper-based analytical devices},\ }\href@noop
  {} {\bibfield  {journal} {\bibinfo  {journal} {Analytical Chemistry}\
  }\textbf {\bibinfo {volume} {82}},\ \bibinfo {pages} {3} (\bibinfo {year}
  {2010})}\BibitemShut {NoStop}%
\bibitem [{\citenamefont {Teh}\ \emph {et~al.}(2008)\citenamefont {Teh},
  \citenamefont {Lin}, \citenamefont {Hung},\ and\ \citenamefont
  {Lee}}]{teh2008droplet}%
  \BibitemOpen
  \bibfield  {author} {\bibinfo {author} {\bibfnamefont {S.-Y.}\ \bibnamefont
  {Teh}}, \bibinfo {author} {\bibfnamefont {R.}~\bibnamefont {Lin}}, \bibinfo
  {author} {\bibfnamefont {L.-H.}\ \bibnamefont {Hung}},\ and\ \bibinfo
  {author} {\bibfnamefont {A.~P.}\ \bibnamefont {Lee}},\ }\bibfield  {title}
  {\bibinfo {title} {Droplet microfluidics},\ }\href@noop {} {\bibfield
  {journal} {\bibinfo  {journal} {Lab on a Chip}\ }\textbf {\bibinfo {volume}
  {8}},\ \bibinfo {pages} {198} (\bibinfo {year} {2008})}\BibitemShut {NoStop}%
\bibitem [{\citenamefont {Ando}\ \emph {et~al.}(2012)\citenamefont {Ando},
  \citenamefont {Liu},\ and\ \citenamefont {Ohl}}]{ando2012homogeneous}%
  \BibitemOpen
  \bibfield  {author} {\bibinfo {author} {\bibfnamefont {K.}~\bibnamefont
  {Ando}}, \bibinfo {author} {\bibfnamefont {A.-Q.}\ \bibnamefont {Liu}},\ and\
  \bibinfo {author} {\bibfnamefont {C.-D.}\ \bibnamefont {Ohl}},\ }\bibfield
  {title} {\bibinfo {title} {Homogeneous nucleation in water in microfluidic
  channels},\ }\href@noop {} {\bibfield  {journal} {\bibinfo  {journal}
  {Physical Review Letters}\ }\textbf {\bibinfo {volume} {109}},\ \bibinfo
  {pages} {044501} (\bibinfo {year} {2012})}\BibitemShut {NoStop}%
\bibitem [{\citenamefont {Nordstrom}\ \emph {et~al.}(2010)\citenamefont
  {Nordstrom}, \citenamefont {Verneuil}, \citenamefont {Arratia}, \citenamefont
  {Basu}, \citenamefont {Zhang}, \citenamefont {Yodh}, \citenamefont {Gollub},\
  and\ \citenamefont {Durian}}]{nordstrom2010microfluidic}%
  \BibitemOpen
  \bibfield  {author} {\bibinfo {author} {\bibfnamefont {K.~N.}\ \bibnamefont
  {Nordstrom}}, \bibinfo {author} {\bibfnamefont {E.}~\bibnamefont {Verneuil}},
  \bibinfo {author} {\bibfnamefont {P.}~\bibnamefont {Arratia}}, \bibinfo
  {author} {\bibfnamefont {A.}~\bibnamefont {Basu}}, \bibinfo {author}
  {\bibfnamefont {Z.}~\bibnamefont {Zhang}}, \bibinfo {author} {\bibfnamefont
  {A.~G.}\ \bibnamefont {Yodh}}, \bibinfo {author} {\bibfnamefont {J.~P.}\
  \bibnamefont {Gollub}},\ and\ \bibinfo {author} {\bibfnamefont {D.~J.}\
  \bibnamefont {Durian}},\ }\bibfield  {title} {\bibinfo {title} {Microfluidic
  rheology of soft colloids above and below jamming},\ }\href@noop {}
  {\bibfield  {journal} {\bibinfo  {journal} {Physical Review Letters}\
  }\textbf {\bibinfo {volume} {105}},\ \bibinfo {pages} {175701} (\bibinfo
  {year} {2010})}\BibitemShut {NoStop}%
\bibitem [{\citenamefont {Collins}\ \emph {et~al.}(2018)\citenamefont
  {Collins}, \citenamefont {O’Rorke}, \citenamefont {Devendran},
  \citenamefont {Ma}, \citenamefont {Han}, \citenamefont {Neild},\ and\
  \citenamefont {Ai}}]{collins2018self}%
  \BibitemOpen
  \bibfield  {author} {\bibinfo {author} {\bibfnamefont {D.~J.}\ \bibnamefont
  {Collins}}, \bibinfo {author} {\bibfnamefont {R.}~\bibnamefont {O’Rorke}},
  \bibinfo {author} {\bibfnamefont {C.}~\bibnamefont {Devendran}}, \bibinfo
  {author} {\bibfnamefont {Z.}~\bibnamefont {Ma}}, \bibinfo {author}
  {\bibfnamefont {J.}~\bibnamefont {Han}}, \bibinfo {author} {\bibfnamefont
  {A.}~\bibnamefont {Neild}},\ and\ \bibinfo {author} {\bibfnamefont
  {Y.}~\bibnamefont {Ai}},\ }\bibfield  {title} {\bibinfo {title} {Self-aligned
  acoustofluidic particle focusing and patterning in microfluidic channels from
  channel-based acoustic waveguides},\ }\href@noop {} {\bibfield  {journal}
  {\bibinfo  {journal} {Physical Review Letters}\ }\textbf {\bibinfo {volume}
  {120}},\ \bibinfo {pages} {074502} (\bibinfo {year} {2018})}\BibitemShut
  {NoStop}%
\bibitem [{\citenamefont {Nagrath}\ \emph {et~al.}(2007)\citenamefont
  {Nagrath}, \citenamefont {Sequist}, \citenamefont {Maheswaran}, \citenamefont
  {Bell}, \citenamefont {Irimia}, \citenamefont {Ulkus}, \citenamefont {Smith},
  \citenamefont {Kwak}, \citenamefont {Digumarthy}, \citenamefont {Muzikansky}
  \emph {et~al.}}]{nagrath2007isolation}%
  \BibitemOpen
  \bibfield  {author} {\bibinfo {author} {\bibfnamefont {S.}~\bibnamefont
  {Nagrath}}, \bibinfo {author} {\bibfnamefont {L.~V.}\ \bibnamefont
  {Sequist}}, \bibinfo {author} {\bibfnamefont {S.}~\bibnamefont {Maheswaran}},
  \bibinfo {author} {\bibfnamefont {D.~W.}\ \bibnamefont {Bell}}, \bibinfo
  {author} {\bibfnamefont {D.}~\bibnamefont {Irimia}}, \bibinfo {author}
  {\bibfnamefont {L.}~\bibnamefont {Ulkus}}, \bibinfo {author} {\bibfnamefont
  {M.~R.}\ \bibnamefont {Smith}}, \bibinfo {author} {\bibfnamefont {E.~L.}\
  \bibnamefont {Kwak}}, \bibinfo {author} {\bibfnamefont {S.}~\bibnamefont
  {Digumarthy}}, \bibinfo {author} {\bibfnamefont {A.}~\bibnamefont
  {Muzikansky}}, \emph {et~al.},\ }\bibfield  {title} {\bibinfo {title}
  {Isolation of rare circulating tumour cells in cancer patients by microchip
  technology},\ }\href@noop {} {\bibfield  {journal} {\bibinfo  {journal}
  {Nature}\ }\textbf {\bibinfo {volume} {450}},\ \bibinfo {pages} {1235}
  (\bibinfo {year} {2007})}\BibitemShut {NoStop}%
\bibitem [{\citenamefont {Burg}\ \emph {et~al.}(2007)\citenamefont {Burg},
  \citenamefont {Godin}, \citenamefont {Knudsen}, \citenamefont {Shen},
  \citenamefont {Carlson}, \citenamefont {Foster}, \citenamefont {Babcock},\
  and\ \citenamefont {Manalis}}]{burg2007weighing}%
  \BibitemOpen
  \bibfield  {author} {\bibinfo {author} {\bibfnamefont {T.~P.}\ \bibnamefont
  {Burg}}, \bibinfo {author} {\bibfnamefont {M.}~\bibnamefont {Godin}},
  \bibinfo {author} {\bibfnamefont {S.~M.}\ \bibnamefont {Knudsen}}, \bibinfo
  {author} {\bibfnamefont {W.}~\bibnamefont {Shen}}, \bibinfo {author}
  {\bibfnamefont {G.}~\bibnamefont {Carlson}}, \bibinfo {author} {\bibfnamefont
  {J.~S.}\ \bibnamefont {Foster}}, \bibinfo {author} {\bibfnamefont
  {K.}~\bibnamefont {Babcock}},\ and\ \bibinfo {author} {\bibfnamefont {S.~R.}\
  \bibnamefont {Manalis}},\ }\bibfield  {title} {\bibinfo {title} {Weighing of
  biomolecules, single cells and single nanoparticles in fluid},\ }\href@noop
  {} {\bibfield  {journal} {\bibinfo  {journal} {Nature}\ }\textbf {\bibinfo
  {volume} {446}},\ \bibinfo {pages} {1066} (\bibinfo {year}
  {2007})}\BibitemShut {NoStop}%
\bibitem [{\citenamefont {Locatelli}\ \emph {et~al.}(2016)\citenamefont
  {Locatelli}, \citenamefont {Pierno}, \citenamefont {Baldovin}, \citenamefont
  {Orlandini}, \citenamefont {Tan},\ and\ \citenamefont
  {Pagliara}}]{locatelli2016single}%
  \BibitemOpen
  \bibfield  {author} {\bibinfo {author} {\bibfnamefont {E.}~\bibnamefont
  {Locatelli}}, \bibinfo {author} {\bibfnamefont {M.}~\bibnamefont {Pierno}},
  \bibinfo {author} {\bibfnamefont {F.}~\bibnamefont {Baldovin}}, \bibinfo
  {author} {\bibfnamefont {E.}~\bibnamefont {Orlandini}}, \bibinfo {author}
  {\bibfnamefont {Y.}~\bibnamefont {Tan}},\ and\ \bibinfo {author}
  {\bibfnamefont {S.}~\bibnamefont {Pagliara}},\ }\bibfield  {title} {\bibinfo
  {title} {Single-file escape of colloidal particles from microfluidic
  channels},\ }\href@noop {} {\bibfield  {journal} {\bibinfo  {journal}
  {Physical Review Letters}\ }\textbf {\bibinfo {volume} {117}},\ \bibinfo
  {pages} {038001} (\bibinfo {year} {2016})}\BibitemShut {NoStop}%
\bibitem [{\citenamefont {Dittrich}\ and\ \citenamefont
  {Manz}(2005)}]{dittrich2005single}%
  \BibitemOpen
  \bibfield  {author} {\bibinfo {author} {\bibfnamefont {P.~S.}\ \bibnamefont
  {Dittrich}}\ and\ \bibinfo {author} {\bibfnamefont {A.}~\bibnamefont
  {Manz}},\ }\bibfield  {title} {\bibinfo {title} {Single-molecule fluorescence
  detection in microfluidic channels—the holy grail in $\mu$\textsc{TAS}?},\
  }\href@noop {} {\bibfield  {journal} {\bibinfo  {journal} {Analytical and
  bioanalytical chemistry}\ }\textbf {\bibinfo {volume} {382}},\ \bibinfo
  {pages} {1771} (\bibinfo {year} {2005})}\BibitemShut {NoStop}%
\bibitem [{\citenamefont {Christov}\ \emph {et~al.}(2011)\citenamefont
  {Christov}, \citenamefont {Lueptow},\ and\ \citenamefont
  {Ottino}}]{christov2011stretching}%
  \BibitemOpen
  \bibfield  {author} {\bibinfo {author} {\bibfnamefont {I.~C.}\ \bibnamefont
  {Christov}}, \bibinfo {author} {\bibfnamefont {R.~M.}\ \bibnamefont
  {Lueptow}},\ and\ \bibinfo {author} {\bibfnamefont {J.~M.}\ \bibnamefont
  {Ottino}},\ }\bibfield  {title} {\bibinfo {title} {Stretching and folding
  versus cutting and shuffling: An illustrated perspective on mixing and
  deformations of continua},\ }\href@noop {} {\bibfield  {journal} {\bibinfo
  {journal} {American Journal of Physics}\ }\textbf {\bibinfo {volume} {79}},\
  \bibinfo {pages} {359} (\bibinfo {year} {2011})}\BibitemShut {NoStop}%
\bibitem [{\citenamefont {Wiggins}\ and\ \citenamefont
  {Ottino}(2004)}]{wiggins2004foundations}%
  \BibitemOpen
  \bibfield  {author} {\bibinfo {author} {\bibfnamefont {S.}~\bibnamefont
  {Wiggins}}\ and\ \bibinfo {author} {\bibfnamefont {J.~M.}\ \bibnamefont
  {Ottino}},\ }\bibfield  {title} {\bibinfo {title} {Foundations of chaotic
  mixing},\ }\href@noop {} {\bibfield  {journal} {\bibinfo  {journal}
  {Philosophical Transactions of the Royal Society of London. Series A:
  Mathematical, Physical and Engineering Sciences}\ }\textbf {\bibinfo {volume}
  {362}},\ \bibinfo {pages} {937} (\bibinfo {year} {2004})}\BibitemShut
  {NoStop}%
\bibitem [{\citenamefont {Aref}(1984)}]{aref1984stirring}%
  \BibitemOpen
  \bibfield  {author} {\bibinfo {author} {\bibfnamefont {H.}~\bibnamefont
  {Aref}},\ }\bibfield  {title} {\bibinfo {title} {Stirring by chaotic
  advection},\ }\href@noop {} {\bibfield  {journal} {\bibinfo  {journal}
  {Journal of Fluid Mechanics}\ }\textbf {\bibinfo {volume} {143}},\ \bibinfo
  {pages} {1} (\bibinfo {year} {1984})}\BibitemShut {NoStop}%
\bibitem [{\citenamefont {Stroock}\ \emph {et~al.}(2002)\citenamefont
  {Stroock}, \citenamefont {Dertinger}, \citenamefont {Ajdari}, \citenamefont
  {Mezi{\'c}}, \citenamefont {Stone},\ and\ \citenamefont
  {Whitesides}}]{stroock2002chaotic}%
  \BibitemOpen
  \bibfield  {author} {\bibinfo {author} {\bibfnamefont {A.~D.}\ \bibnamefont
  {Stroock}}, \bibinfo {author} {\bibfnamefont {S.~K.}\ \bibnamefont
  {Dertinger}}, \bibinfo {author} {\bibfnamefont {A.}~\bibnamefont {Ajdari}},
  \bibinfo {author} {\bibfnamefont {I.}~\bibnamefont {Mezi{\'c}}}, \bibinfo
  {author} {\bibfnamefont {H.~A.}\ \bibnamefont {Stone}},\ and\ \bibinfo
  {author} {\bibfnamefont {G.~M.}\ \bibnamefont {Whitesides}},\ }\bibfield
  {title} {\bibinfo {title} {Chaotic mixer for microchannels},\ }\href@noop {}
  {\bibfield  {journal} {\bibinfo  {journal} {Science}\ }\textbf {\bibinfo
  {volume} {295}},\ \bibinfo {pages} {647} (\bibinfo {year}
  {2002})}\BibitemShut {NoStop}%
\bibitem [{\citenamefont {Khodaparast}\ \emph {et~al.}(2017)\citenamefont
  {Khodaparast}, \citenamefont {Boulogne}, \citenamefont {Poulard},\ and\
  \citenamefont {Stone}}]{khodaparast2017water}%
  \BibitemOpen
  \bibfield  {author} {\bibinfo {author} {\bibfnamefont {S.}~\bibnamefont
  {Khodaparast}}, \bibinfo {author} {\bibfnamefont {F.}~\bibnamefont
  {Boulogne}}, \bibinfo {author} {\bibfnamefont {C.}~\bibnamefont {Poulard}},\
  and\ \bibinfo {author} {\bibfnamefont {H.~A.}\ \bibnamefont {Stone}},\
  }\bibfield  {title} {\bibinfo {title} {Water-based peeling of thin
  hydrophobic films},\ }\href@noop {} {\bibfield  {journal} {\bibinfo
  {journal} {Physical Review Letters}\ }\textbf {\bibinfo {volume} {119}},\
  \bibinfo {pages} {154502} (\bibinfo {year} {2017})}\BibitemShut {NoStop}%
\bibitem [{\citenamefont {Xia}\ and\ \citenamefont
  {Whitesides}(1998)}]{xia1998soft}%
  \BibitemOpen
  \bibfield  {author} {\bibinfo {author} {\bibfnamefont {Y.}~\bibnamefont
  {Xia}}\ and\ \bibinfo {author} {\bibfnamefont {G.~M.}\ \bibnamefont
  {Whitesides}},\ }\bibfield  {title} {\bibinfo {title} {Soft lithography},\
  }\href@noop {} {\bibfield  {journal} {\bibinfo  {journal} {Annual review of
  materials science}\ }\textbf {\bibinfo {volume} {28}},\ \bibinfo {pages}
  {153} (\bibinfo {year} {1998})}\BibitemShut {NoStop}%
\bibitem [{\citenamefont {Barnes}\ \emph {et~al.}(2021)\citenamefont {Barnes},
  \citenamefont {Ouro-Koura}, \citenamefont {Derickson}, \citenamefont
  {Lebarty}, \citenamefont {Omidokun}, \citenamefont {Bane}, \citenamefont
  {Suleiman}, \citenamefont {Omagamre}, \citenamefont {Fotouhi}, \citenamefont
  {Ogunmolasuyi} \emph {et~al.}}]{barnes2021plasma}%
  \BibitemOpen
  \bibfield  {author} {\bibinfo {author} {\bibfnamefont {B.~K.}\ \bibnamefont
  {Barnes}}, \bibinfo {author} {\bibfnamefont {H.}~\bibnamefont {Ouro-Koura}},
  \bibinfo {author} {\bibfnamefont {J.}~\bibnamefont {Derickson}}, \bibinfo
  {author} {\bibfnamefont {S.}~\bibnamefont {Lebarty}}, \bibinfo {author}
  {\bibfnamefont {J.}~\bibnamefont {Omidokun}}, \bibinfo {author}
  {\bibfnamefont {N.}~\bibnamefont {Bane}}, \bibinfo {author} {\bibfnamefont
  {O.}~\bibnamefont {Suleiman}}, \bibinfo {author} {\bibfnamefont
  {E.}~\bibnamefont {Omagamre}}, \bibinfo {author} {\bibfnamefont {M.~J.}\
  \bibnamefont {Fotouhi}}, \bibinfo {author} {\bibfnamefont {A.}~\bibnamefont
  {Ogunmolasuyi}}, \emph {et~al.},\ }\bibfield  {title} {\bibinfo {title}
  {Plasma generation by household microwave oven for surface modification and
  other emerging applications},\ }\href@noop {} {\bibfield  {journal} {\bibinfo
   {journal} {American Journal of Physics}\ }\textbf {\bibinfo {volume} {89}},\
  \bibinfo {pages} {372} (\bibinfo {year} {2021})}\BibitemShut {NoStop}%
\bibitem [{\citenamefont {Bhattacharya}\ \emph {et~al.}(2005)\citenamefont
  {Bhattacharya}, \citenamefont {Datta}, \citenamefont {Berg},\ and\
  \citenamefont {Gangopadhyay}}]{bhattacharya2005studies}%
  \BibitemOpen
  \bibfield  {author} {\bibinfo {author} {\bibfnamefont {S.}~\bibnamefont
  {Bhattacharya}}, \bibinfo {author} {\bibfnamefont {A.}~\bibnamefont {Datta}},
  \bibinfo {author} {\bibfnamefont {J.~M.}\ \bibnamefont {Berg}},\ and\
  \bibinfo {author} {\bibfnamefont {S.}~\bibnamefont {Gangopadhyay}},\
  }\bibfield  {title} {\bibinfo {title} {Studies on surface wettability of poly
  (dimethyl) siloxane (pdms) and glass under oxygen-plasma treatment and
  correlation with bond strength},\ }\href@noop {} {\bibfield  {journal}
  {\bibinfo  {journal} {Journal of Microelectromechanical Systems}\ }\textbf
  {\bibinfo {volume} {14}},\ \bibinfo {pages} {590} (\bibinfo {year}
  {2005})}\BibitemShut {NoStop}%
\bibitem [{\citenamefont {Barkla}\ and\ \citenamefont
  {Auchterlonie}(1971)}]{barkla1971magnus}%
  \BibitemOpen
  \bibfield  {author} {\bibinfo {author} {\bibfnamefont {H.}~\bibnamefont
  {Barkla}}\ and\ \bibinfo {author} {\bibfnamefont {L.}~\bibnamefont
  {Auchterlonie}},\ }\bibfield  {title} {\bibinfo {title} {The magnus or robins
  effect on rotating spheres},\ }\href@noop {} {\bibfield  {journal} {\bibinfo
  {journal} {Journal of Fluid Mechanics}\ }\textbf {\bibinfo {volume} {47}},\
  \bibinfo {pages} {437} (\bibinfo {year} {1971})}\BibitemShut {NoStop}%
\bibitem [{\citenamefont {Ottino}(1989)}]{ottino1989kinematics}%
  \BibitemOpen
  \bibfield  {author} {\bibinfo {author} {\bibfnamefont {J.~M.}\ \bibnamefont
  {Ottino}},\ }\href@noop {} {\emph {\bibinfo {title} {The kinematics of
  mixing: stretching, chaos, and transport}}},\ Vol.~\bibinfo {volume} {3}\
  (\bibinfo  {publisher} {Cambridge university press},\ \bibinfo {year}
  {1989})\BibitemShut {NoStop}%
\bibitem [{\citenamefont {Jones}\ \emph {et~al.}(1989)\citenamefont {Jones},
  \citenamefont {Thomas},\ and\ \citenamefont {Aref}}]{jones1989chaotic}%
  \BibitemOpen
  \bibfield  {author} {\bibinfo {author} {\bibfnamefont {S.~W.}\ \bibnamefont
  {Jones}}, \bibinfo {author} {\bibfnamefont {O.~M.}\ \bibnamefont {Thomas}},\
  and\ \bibinfo {author} {\bibfnamefont {H.}~\bibnamefont {Aref}},\ }\bibfield
  {title} {\bibinfo {title} {Chaotic advection by laminar flow in a twisted
  pipe},\ }\href@noop {} {\bibfield  {journal} {\bibinfo  {journal} {Journal of
  Fluid Mechanics}\ }\textbf {\bibinfo {volume} {209}},\ \bibinfo {pages} {335}
  (\bibinfo {year} {1989})}\BibitemShut {NoStop}%
\end{thebibliography}%

\end{document}